\documentclass[preprint]{aastex}
\usepackage{epsf}

\tighten

\slugcomment{Accepted for Publication in ApJ.}

\newcommand {\cps} {$\,$cts$\,$s$^{-1}$}

\newcommand {\nh}  {N$_{\rm H}$}
\newcommand {\sqig} {$\sim$}
\newcommand {\asca} {{\it ASCA}}
\newcommand {\rosat} {{\it ROSAT}}

\newcommand {\ein} {{\it Einstein}}

\newcommand {\uw} {MS~1603.6+2600}

\def\plotoneb#1#2{\centering \leavevmode
\epsfxsize=#2 \epsfbox{#1}}

\shorttitle{MS~1603.6+2600: a Dipping LMXB in the Halo?}
\shortauthors{Mukai et al.}

\begin{document}

\title{ASCA Observation of MS~1603.6+2600 (=UW~Coronae Borealis): \\
a Dipping Low-Mass X-ray Binary in the Outer Halo?}

\author{Koji Mukai\altaffilmark{1}, Alan P. Smale\altaffilmark{1},
Caroline K. Stahle}
\affil{Code 662, NASA/Goddard Space Flight Center, Greenbelt, MD 20771.}
\author{Eric M. Schlegel}
\affil{Harvard Smithsonian Center for Astrophysics, 60 Garden Street,
Cambridge, MA 02138.}
\author{Rudy Wijnands\altaffilmark{2}}
\affil{Center for Space Research, Massachusetts Institute for
Technology, 77 Massachusetts Avenue, Cambridge, MA 02139.}

\altaffiltext{1}{Also Universities Space Research Association}
\altaffiltext{2}{Chandra Fellow}

\begin{abstract}
\uw\ is a high-latitude X-ray binary with a 111 min orbital period,
thought to be either an unusual cataclysmic variable or an unusual
low-mass X-ray binary.  In an \asca\ observation in 1997 August,
we find a burst, whose light curve suggests a Type I (thermonuclear
flash) origin.  We also find an orbital X-ray modulation in \uw,
which is likely to be periodic dips, presumably due to azimuthal
structure in the accretion disk.  Both are consistent with this
system being a normal low-mass X-ray binary harboring a neutron star,
but at a great distance.  We tentatively suggest
that \uw\ is located in the outer halo of the Milky Way, perhaps associated
with the globular cluster Palomar 14, 11$^\circ$ away from \uw\ on the
sky at an estimated distance of 73.8 kpc.
\end{abstract}


\section{Introduction}

\uw\ was first discovered in the course of the Extended Medium Sensitivity
Survey \citep{g90}.  With only 51 detected photons in a 2112 s exposure,
this detection was useful only in providing the position and a flux
estimate of $\sim 10^{-12}$ ergs\,cm$^{-2}$\,s$^{-1}$ (0.3--3.5 keV).
\citet{m90} identified the optical counterpart, subsequently designated
UW~Coronae~Borealis, a V \sqig 19.7th mag object.  It was shown to be
an eclipsing binary: measurements of 10 eclipse timings over a 3-day
period has enabled \citet{m90} to derive a 111.0 min orbital period.
This immediately establishes \uw\ as a low-mass, compact binary.  To
fit in such a short period binary, the secondary must be a low mass star;
and, to be a significant X-ray source, the primary must be an accreting
compact object --- a white dwarf, which would make the system a cataclysmic
variable (CV), a neutron star, or perhaps a black hole, either of which
would make it a low mass X-ray binary (LMXB).  Given the high galactic
latitude of this object ($b^{\rm II}\sim47^\circ$), it has to be at a
large distance ($d \gg 10$ kpc) to be a normal, bright LMXB.  On the
other hand, the optical spectrum and the high X-ray to optical flux ratio
(f$_X$/f$_{opt}$ \sqig 15) are unlike those of normal CVs.

Further optical observations \citep{v93,h98} have revealed highly
variable orbital light curves.  Report of a 112.5 min period,
obtained by \citet{v93} from Fourier analysis, would make it a system
with multiple periods, if confirmed.  A \rosat\ PSPC observation performed
on 1991 August 26 \citep{h98} did not resolve the nature of \uw.
\citet{h98} discussed the possibility that \uw\ may be an LMXB with
an accretion disk corona (ADC), in which we only observe a small
fraction of X-rays scattered in the ADC surrounding the inner disk
and the neutron star.  They also considered the possibility that it
might be a transient LMXB in quiescence.  These scenarios are motivated,
in part, by the desire to keep \uw\ within a few kiloparsecs of the
Galactic plane.

We have obtained an \asca\ observation of \uw\ in an effort to clarify
the nature of this object (see also our preliminary report in \citet{m99}). 
We describe the observation in \S 2, the results in \S3, and discuss the
implications in \S4.

\section{Observation}

\uw\ was observed with \asca\ \citep{t94} from 1997 August 30 at 20:25 UT
to 1997 August 31 at 13:52 UT.  Of the 4 instruments on-board \asca, the
2 GIS's were operated in the standard PH mode.  For the GIS data,
we have selected intervals when the satellite was outside the South
Atlantic Anomaly (SAA), the attitude control was stable and within
0.02$^\circ$ of the target, and the line-of-sight was $>5^\circ$
above the Earth limb.  We have also applied a standard selection
expression combining monitor count rates and geomagnetic cut-off
rigidity (COR) to exclude time intervals of high particle background,
obtaining \sqig 25 ksec of good on-source data.  We extracted the source
spectra and light curves using 6 arcmin radius circular extraction regions,
and used \sqig 13 arcmin radius regions around the center of the detectors,
excluding regions within \sqig 7.5 arcmin of the source, as background
(no sources were detected in the background regions both for GIS and for
SIS).  The net count rates were \sqig 0.06 \cps\ in GIS-2 and \sqig 0.07
\cps\ in GIS-3.

For the 2 SIS instruments, we have applied the same SAA, attitude, and
elevation constraints, and further excluded data taken within 10$^\circ$
of the bright Earth limb.  We have selected low background time intervals
using COR and PIXEL monitor count criteria, and also excluded data taken
within 32 seconds of day/night and SAA transitions, when the on-board
dark frame calculations are suspect.  These resulted in \sqig 25 ksec
of useful SIS data.  We used an extraction region of \sqig 4.5 arcmin
radius for the SIS-0, using the entire chip, excluding regions
within \sqig 5.5 arcmin of the source, as background, with a net rate 
of \sqig 0.12 \cps.  For SIS-1, we were forced to use a smaller (\sqig 3.5
arcmin) source region to stay entirely within the active chip,
obtaining a net rate of \sqig 0.08 \cps.

For spectral analysis, we combined data from like instruments to form
a GIS spectrum and an SIS spectrum.  The latter is affected by the secular
change in SIS performance at low energies; this can be phenomenologically
characterized as a spurious excess absorption of order \nh\ =
$1.0 \times 10^{21}$ cm$^{-2}$.  For the light curve analysis, we combined
data from all 4 instruments, both using the entire \asca\ passband,
and in two sub-bands, one below 2 keV (0.48--2.0 for SIS and 0.7--2.0 keV
for GIS) and the other above 2 keV (up to 10 keV).

\section{Results}

\subsection{Detection of a burst}

\begin{figure}[t]
\begin{center}
\plotoneb{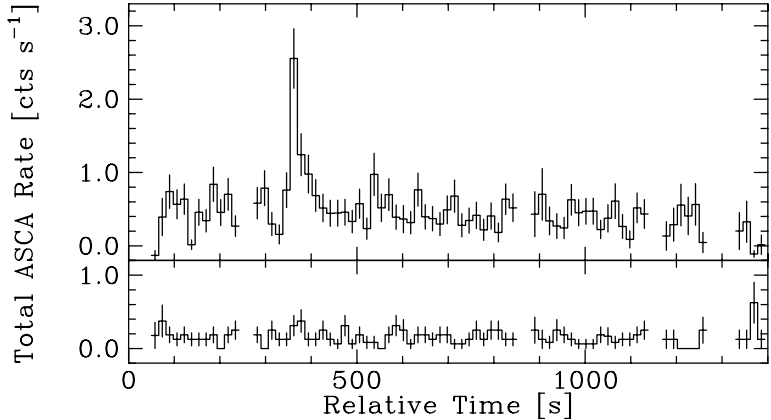}{10.5 cm}
\caption{(Upper panel) \asca\ light curve of the burst, in 16 s bins. 
We show the total count rate for 2 SIS (0.48--10 keV) and 2 GIS
(0.7--10 keV) instruments.  (Lower panel) The background light curve
during the same interval.}
\end{center}
\end{figure}

We present a 16-s bin light curve of a striking, flare-like event
in Figure 1.  This event, detected at 21:45:50 UT on 1997 August 30,
is seen in all 4 individual source light curves, and in none of the
4 background light curves.  We are therefore confident that this is
a real event, presumably due to \uw.  The rise is rapid, and the decay
lasts for about 1 minute before blending into the (noisy) persistent
emission.  The peak count rate ($\sim$ 2.6 \cps\ in 4 detectors) is 
8 times the persistent emission.

We have attempted a more detailed analysis of this event, but this
has not been fruitful because there are only \sqig 60 net counts
from this event.  In the remainder of this paper, we will refer to
this event as a ``burst.''  The resemblance with a Type I X-ray burst
(i.e., a thermonuclear runaway on the surface of an accreting neutron
star; see \citet{l95} for a review) is suggestive, but not conclusive.

\subsection{Orbital modulation}

\begin{figure}[t]
\begin{center}
\plotoneb{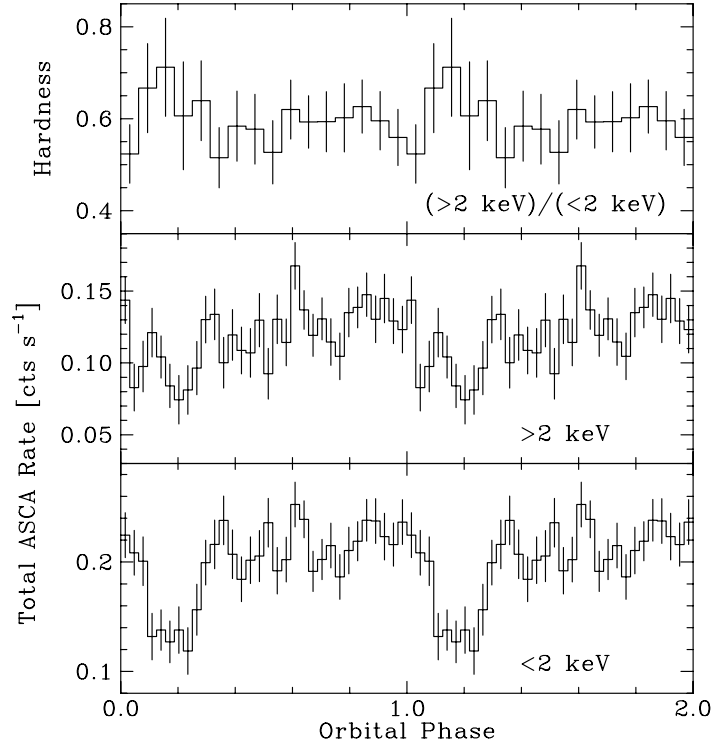}{9.5 cm}
\caption{The \asca\ light curves of \uw\ above and below 2 keV, folded on the
orbital ephemeris of \citet{m90}, in 32 bins per cycle, plotted twice
for clarity.  A hardness ratio plot is shown on the top panel in 16 bins
per cycle.}
\end{center}
\end{figure}

Next, we removed the burst from our light curves and performed
Fourier and folding analyses.  We find a strong periodicity at 116$\pm$6
min, the large uncertainty being due to the limited duration of the
observation and data gaps.
We interpret this signal as reflecting the orbital modulation, and proceed
below by folding the data on the orbital ephemeris of \citet{m90}.  Note,
however, that this ephemeris does not define phase 0.0 accurately.
No other periodicities are apparent in the data in the 200--20,000 s range.

We present the folded orbital light curves, in high ($>$2 keV) and
low ($<$2 keV) energy bands, as well as the hardness ratio, in Figure 2.
A clear orbital minimum, lasting \sqig 15\% of the orbit and reaching
\sqig 60\% of the flux level outside the minimum, is seen in the
low energy light curve.  It is less clear in the high energy curve,
which is reflected in the slight increase in the hardness ratio
at these phase intervals.  Inspection of individual orbital cycles
shows that the minimum is always present during the same phase in
broad outline, though their details vary.  However, given the
statistics, we cannot rule out statistical fluctuation as the cause
of the cycle-to-cycle variations.  Outside the minimum, the light
curves are flat, with apparently random fluctuations superimposed.

\begin{figure}[t]
\begin{center}
\plotoneb{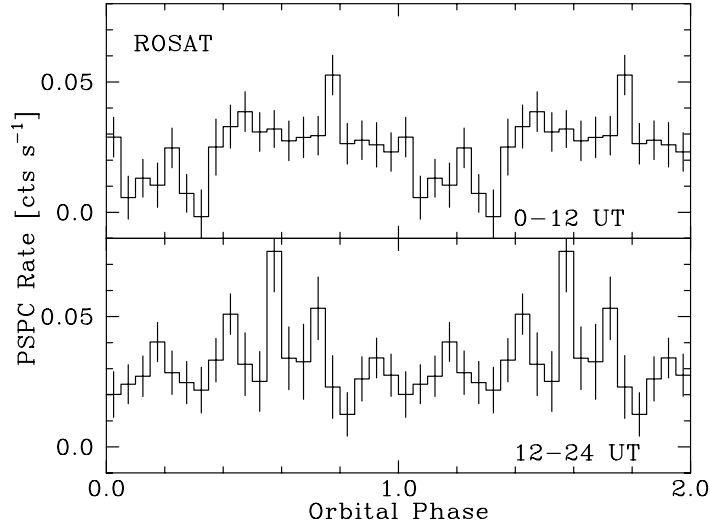}{9.5 cm}
\caption{The archival \rosat\ PSPC light curve of \uw, obtained on 1991
August 26.  The data have been divided into two sections, and folded on
the orbital period with 20 bins.}
\end{center}
\end{figure}

This behavior is rather different from that reported by \citet{h98}
in their 1991 \rosat\ PSPC observation.  To investigate
this further, we have retrieved the \rosat\ data from the HEASARC and
re-extracted the light curves.  At first glance, we merely confirmed
the results of \citet{h98}.  However, after dividing the \rosat\ data
into two halves of roughly equal durations (\sqig 12 hrs), we find a pattern
of orbital minimum for the first half of the \rosat\ observation (Figure 3).
This minimum is absent during the second half; instead, the folded
light curve shows a flaring-like behavior.

\subsection{Spectral analysis}

\begin{figure}[t]
\begin{center}
\plotoneb{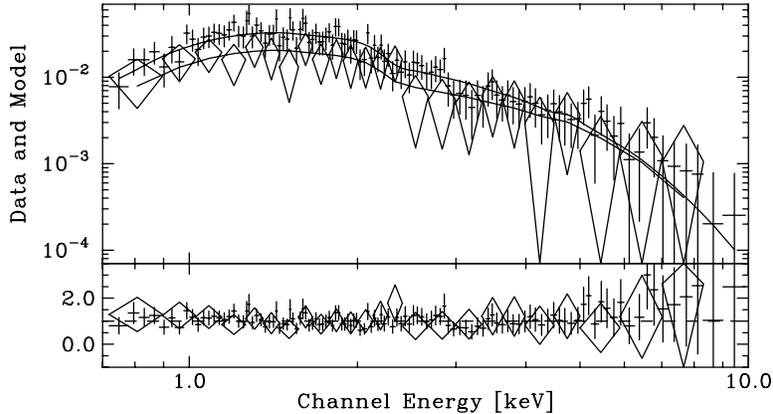}{10.5 cm}
\caption{\asca\ GIS spectra of \uw\ during orbital maximum (pluses) and minimum
(diamonds).  The top panel shows the data and the model (see text for details),
while the bottom panel shows the data to model ratio.}
\end{center}
\end{figure}

Despite the improvement over the \rosat\ PSPC spectrum, we have not been able
to draw a definite conclusion from the average \asca\ spectrum of \uw.
It can be fit with a power law, a bremsstrahlung, or a Comptonization model:
discrimination among these continuum models proved impossible because high
signal-to-noise ratio was not achieved over a sufficiently wide bandpass.
No discrete features are obvious.  Regardless of the continuum model, the
0.7--10 keV flux of \uw\ during the \asca\ observation was
\sqig $4.0 \times 10^{-12}$ ergs\,cm$^{-2}$\,s$^{-1}$
(\sqig $4.8 \times 10^{34}$ [d/10 kpc]$^2$ ergs\,s$^{-1}$).
Ignoring the difference in spectral shapes, the \sqig 8 fold increase
in count rate at the peak of the burst implies a peak luminosity of
\sqig $3.8 \times 10^{35}$ [d/10 kpc]$^2$ ergs\,s$^{-1}$).

The average \asca\ flux value indicates a significant long-term
variability in \uw.  The \ein\ IPC 0.3-3.5 keV flux of
1.14 $\times 10^{-12}$ ergs\,cm$^{-2}$\,s$^{-1}$ \citep{m90} is
roughly a factor of 4 lower than the \asca\ value, although both
the bandpass and the assumed spectral shape are different.  Back
prediction of \ein\ count rate based on \asca\ flux and the best-fit
bremsstrahlung model also points to a factor of \sqig 4 change in flux.
Similarly, \uw\ appears to have been fainter by a factor of \sqig 7
during the \rosat\ observation, compared to 1997 August.

We also compared the spectrum during the orbital minimum to that of the
maximum, by simultaneously fitting the two using the same bremsstrahlung
model (identical kT and normalization).  For the maximum,
we assumed no absorption, since the interstellar column, of the
order of $10^{20}$ cm$^{-2}$ according to the \rosat\ PSPC spectral
fit \citep{h98}, is too small to be detectable with \asca.
For the minimum spectrum, we have added an absorption component.
Either a simple absorber of \nh\ = $4.6 \times 10^{21}$ cm$^{-2}$
($\chi^2$=70.6 for 140 PHA bins and 4 degrees of freedom)
or a partial covering absorber of \nh\ = $1.3 \times 10^{23}$ cm$^{-2}$
and a covering fraction of 37\% ($\chi^2$=64.8 for 140 PHA bins and
5 degrees of freedom) can be used to fit both spectra.  The latter
is shown for the GIS data in Figure 4.

\section{Discussion: the Nature and the Location of \uw}

We have observed an orbital minimum in the \asca\ light curve
which is broad and partial.  Our re-analysis of the \rosat\ PSPC
light curve shows that a similar minimum was present during the
first half of the 1991 observation, but disappeared during the
second half.

In the ADC model that we favored in our earlier report \citep{m99},
this minimum is caused by a partial eclipse of an extended X-ray source,
whose size is a significant fraction of the binary separation.
An ADC source is an LMXB seen at a high inclination angle.
The central X-ray emitting region is blocked from our direct view by the
accretion disk.  A fraction (1--10\%) is scattered into our line of
site by the ADC above the accretion disk.  Although this interpretation
has several attractive properties, \uw\ does not show the smooth,
quasi-sinusoidal modulation which is seen in the prototype ADC source,
X1822$-$372 \citep{h89}.  The orbital light curves of ADC sources
generally appear to be stable from one cycle to the next, while the
\rosat\ data show \uw\ to be otherwise.  In addition, the spectrum of
X1822$-$372 is softer during the partial eclipse, which is the reverse
of what we observe in \uw.  Therefore, the behavior of \uw\ differs
in detail from those of well-established ADC sources such as X1822$-$372.
Moreover, the ingress into the orbital minimum in \uw\ (Figure 2) takes
place in less than one phase bin (1/32th of the orbital cycle).  It is
difficult to construct a geometry of an extended X-ray source that leads to
a broad, partial eclipse yet with such a relatively rapid ingress.

On the other hand, the behavior of \uw\ is similar to that of dipping
LMXBs such as XB~1916$-$053 \citep{h01}.  In dippers, we believe that
the central X-ray source (the accretion disk, the neutron star, or both) is
occulted by azimuthal structure of the accretion disk.  The preferential
location of the structure in binary frame defines the envelope of
phases during which dips occur.  The dip durations and depth change
from epoch to epoch, sometimes within a single observation.  Spectrally,
dips are deeper at low energies, although simple absorption models often
do not fit the data, instead requiring a partial covering absorber.
  In all these respects, the \asca\ data of \uw\ are consistent with
the dipper interpretation.  We therefore favor this interpretation
over the ADC model.

As a dipper, \uw\ must possess an accretion disk.  This immediately
excludes the AM Her-type magnetic CVs (in which the magnetic field
is so strong as to prevent the formation of a disk) from consideration.
It is still possible that \uw\ is a dipping CV with an accretion disk.
For example, U Gem, the prototype dwarf nova, is a dipper \citep{s96}.
However, all the objections raised to date remain valid, most notably
that the f$_X$/f$_{opt}$ ratio is too high: U Gem is \sqig 1,000 times
brighter than \uw\ in the optical while only \sqig 4 times brighter in
the \asca\ band.  Magnetic CVs of the intermediate polar type, which has
partial accretion disk, also show X-ray dips \citep{h93}.  However,
their defining characteristic is a strong spin modulation in the X-rays,
which we do not see.  Thus, if \uw\ is a CV, it must be an unusual
example.

Moreover, a burst-like event like the one we have seen in \uw\ has
never been reported in any X-ray observations of CVs.  We therefore
assume, as a working hypothesis, that \uw\ is an LMXB containing a
neutron star and the burst is a Type I event.  This suggest \uw\ is
at a great distance, since the typical persistent luminosity of a
dipper is a few times $10^{36}$ ergs\,s$^{-1}$
and the typical burst peak luminosity is a few times $10^{37}$ ergs\,s$^{-1}$.
Given the observed flux values, this would require $d \gg 10$ kpc.

Is this a serious weakness of this model?  Is there anything in the
Galactic halo out to, say, $d$ \sqig 50 kpc?  To answer these questions,
we have conducted a simple search for globular clusters near the direction
of \uw.  The nearest match was found to be Palomar 14 \citep{s97},
\sqig 11$^\circ$ away on the sky at an estimated heliocentric distance
of 73.8 kpc.

This suggests possible, though speculative, scenarios as to how
\uw\ might have formed in the outer halo.  First, it may have formed
in Palomar 14 itself and escaped.  Of the \sqig 150 Galactic globular
clusters, 12 harbor a luminous X-ray binary \citep{d00}.  The same
stellar encounters that form these LMXBs eject a significant number
of resultant binaries. One calculation \citep{p97} suggest the fraction
of the escaped binaries may be about a third of those that have survived
in the globular cluster, thus we expect several ``non-globular-cluster''
LMXBs to have escaped from the parent cluster.  Similar escapees from
a cluster much closer to the Galactic center would have been assimilated
into the Galactic bulge, and its origin would have become obscured.
If \uw\ was ejected from Palomar 14 at a velocity of 10 km\,s$^{-1}$,
it would have taken it \sqig 1 Gyr to move 10 kpc, which is the minimum
distance implied by the current angular separation of 11$^\circ$.  Thus
it appears possible to have a reasonable combination of space velocity
and lifetime of \uw\ to explain its current location, for an assumed
origin in Palomar 14.

Another speculative scenario is related to the possible origin of
Palomar 14, and many other outer halo globular clusters, in dwarf
elliptical satellites of the Milky Way \citep{v00}.  Several
outer halo clusters are associated with the Sagittarius dwarf.
Similarly, Palomar 14 may be associated with an undiscovered extant
satellite, or one that has been tidally disrupted.  In either case,
\uw\ could be a member of this hypothetical satellite galaxy.

If we assume a physical association between \uw\ and Palomar 14,
at a distance of $d \sim 70$ kpc, then the average luminosity
during \asca\ observation was \sqig $2.4 \times 10^{36}$ ergs\,s$^{-1}$
and the peak burst luminosity was \sqig $1.9 \times 10^{37}$ ergs\,s$^{-1}$,
both of which are normal for a bursting, dipping LMXB, although still
with an unusually low f$_X$/f$_{opt}$ ratio.  Evolutionary scenario
Ia of \citet{e93}, similar to short period CVs, predicts a similar
luminosity and bursting behavior, and hence fits our outer halo interpretation.
However, this may be a coincidence if \uw\ is indeed formed
in Palomar 14, since globular cluster LMXBs form in a distinct manner
from field LMXBs.

If our interpretation is correct, \uw\ is a close analog of XB~1916$-$053,
including the strong variabilities of the dip morphology.  In fact,
the second optical period obtained via Fourier transform \citep{v93}
may be real, indicating that \uw\ is a permanent superhumper \citep{has01},
another characteristic of XB~1916$-$053.  If so, it would be possible to
estimate the mass ratio of \uw\ using the fractional period excess.

\section{Summary and Future Prospects}

We have presented our \asca\ observation of the unusual high-latitude
X-ray binary, \uw.  We have detected a burst and an orbital modulation
consistent with a dipping behavior, and presented an outer halo LMXB
model as perhaps the least contrived interpretation.

Future X-ray observations can test our interpretation.  In particular,
the detection of other bursts with better sensitivity is necessary to
confirm the Type I interpretation spectroscopically.  Moreover, if we are
lucky enough to catch a radius expansion burst, a reliable distance
and other parameters can be derived.  The ingress and egress to the
orbital minimum should be investigated further, so that we can can
conclusively choose between the dipper and the ADC interpretations.
At the same time, further optical observations should clear up the
issue of possible multiple periodicities.

Pending such observations, we tentatively conclude that \uw\ may
be the most distant Galactic X-ray binary known.  Moreover, this may
be an escaped globular cluster LMXB, something we cannot confidently
observe near the Galactic center, or may be a relic of the hierarchical
formation of the Milky Way halo.

\acknowledgments

This research has made use of data obtained from the
High Energy Astrophysics Science Archive Research Center (HEASARC),
provided by NASA's Goddard Space Flight Center.  RW is supported by
NASA through Chandra Postdoctoral Fellowship grant PF9-10010 awarded
by CXC, which is operated by SAS for NASA under contract NAS8-39073.

\end{document}